\documentclass[12pt,a4paper,final]{iopart}

\usepackage{iopams} 
\expandafter\let\csname equation*\endcsname\relax
\expandafter\let\csname endequation*\endcsname\relax
\usepackage{amsmath, amsthm, amssymb, amsfonts}
\usepackage{graphicx}
\usepackage[breaklinks=true,colorlinks=true,linkcolor=blue,urlcolor=blue,citecolor=blue]{hyperref}
\usepackage{braket}
\newcommand{\ba}{\begin{eqnarray}}
\newcommand{\ea}{\end{eqnarray}}
\newcommand{\ban}{\begin{eqnarray*}}
\newcommand{\ean}{\end{eqnarray*}}

\begin{document}

\title[Y. Wang et al., All the self-testings of the singlet for two binary measurements]{All the self-testings of the singlet for two binary measurements}

\author{Yukun Wang}
\address{State key Laboratory of Networking and Switching Technology, Beijing University of Posts and Telecommunications, Beijing, China 100876}
\address{Centre for Quantum Technologies, National University of Singapore, 3 Science Drive 2, Singapore 117543}

\author{Xingyao Wu}
\address{Centre for Quantum Technologies, National University of Singapore, 3 Science Drive 2, Singapore 117543}

\author{Valerio Scarani}
\address{Centre for Quantum Technologies, National University of Singapore, 3 Science Drive 2, Singapore 117543}
\address{Department of Physics, National University of Singapore, 2 Science Drive 3, Singapore 117542}

\begin{abstract}
Self-testing refers to the possibility of characterizing uniquely (up to local isometries) the state and measurements contained in quantum devices, based only on the observed input-output statistics. Already in the basic case of the two-qubit singlet, self-testing is not unique: the two known criteria (the maximal violation of the CHSH inequality and the Mayers-Yao correlations) are not equivalent. It is unknown how many criteria there are. In this paper, we find the whole set of criteria for the ideal self-testing of a singlet with two measurements and two outcomes on each side: it coincides with all the extremal points of the quantum set that can be obtained by measuring the singlet.
\end{abstract}

\pacs{00.00, 20.00, 42.10}
\vspace{2pc}
\noindent{\it Keywords}: Device-Independent, self-testing,  CHSH and Mayers-Yao.

\section{Introduction}

The idea of using device-independent (DI) approach for secure quantum information processing has been around for almost a decade. In DI approach, the experiments are described from the observed statistics of measurement outcomes. Under the sole assumptions of no-signaling and the validity of quantum theory, it has been possible to assess the performance of quantum devices in quantum key distribution \cite{ref1}, randomness generation \cite{ref2}, entanglement witness and dimension witness \cite{refRN}. The certification of the performance of unknown devices in some specific tasks is already a remarkable feat; but in some cases, one can do even more and certify uniquely the state and the measurements that are present in the devices.

In fact, the possibility of this ``self-testing" in the ideal case is known since the 1990's, when it was proved that the maximal violation of the CHSH Bell inequality \cite{ref4} identifies uniquely the maximally entangled state of two qubits \cite{refSJ,refSD,refBS}, which will be referred to as `singlet' from now on. A few years later, another criterion was given by Mayers and Yao \cite{refMY} (see also \cite{refTF}), for the same state and under the same ideal conditions. The tools developed in the context of DI assessment make it possible to go beyond these ideal and highly specialized statements. In particular, robustness bounds for self-testing singlet with these two criteria were presented \cite{refmm,refth,refjd}.


The observation that motivate this paper is that the two known criteria to self-test the singlet, CHSH and Mayers-Yao, are \textit{not equivalent}: indeed, from the Mayers-Yao correlations one can get at most $CHSH=\sqrt{2}+1<2\sqrt{2}$. The work of Miller and Shi \cite{millershi} suggests that many more self-testing criteria can actually be defined. The fact that one state may have several ``signatures" is not only interesting for the better understanding of self-testing: it may have implication at the moment of actually implementing the certification. Indeed, the two known criteria don't seem to perform equally well: for feasible levels of approximation of the quantum set by semi-definite criteria \cite{refMS,refMN}, the CHSH test was found to be less sensitive to imperfections \cite{refth,refjd}.

The main result of this paper is the characterization of \textit{all} the criteria that self-test the singlet using two measurements with two outcomes on each party. These criteria are given directly in terms of the observed statistics.

\section{Definition of self-testing}

We consider 2 parties, Alice and Bob, each having a device with 2 inputs (``measurements"), and each measurement has 2 outcomes [(2,2,2) scenario]. Alice's inputs and outputs are denoted respectively by $x$ and $a$, Bob's by $y$ and $b$; and we choose the label $a,b = \pm 1$ for the local setting $x,y \in \{0, 1\}$. No assumption on the internal working of  Alice's and Bob's devices, besides the fact that no information about one's input is available to the other's box to produce its outcome. After a large number of queries, Alice and Bob can reconstruct the joint probability distribution $p(a,b|x,y)$. 

We assume that the boxes are correctly described by quantum theory: there exist a state $\rho$ and four measurements $\{\Pi_a^x\}$, $\{\Pi_b^y\}$ such that $p(a,b|x,y)=\Tr(\rho\Pi_a^x\Pi_b^y)$. The no-signaling assumption noted above implies $[\Pi_a^x,\Pi_b^y]=0$. Since the dimension is not fixed, the measurements can be taken as projective without loss of generality. Similarly, since nothing is said about the degrees of freedom on which the measurements act, the purification of the state can be included in the boxes, so the state can be considered pure.

The mapping of classical statistics to quantum system is usually one-to-many. \textit{Self-testing} refers to the cases where a $p(a,b|x,y)$ predicted by quantum theory determines uniquely the state (which in these cases is pure) and the measurements, up to a local isometry $\Phi=\Phi_A\otimes\Phi_B$. Specifically, for the two-qubit singlet $|\Phi^{+}\rangle=\frac{|00\rangle+|11\rangle}{\sqrt{2}}$, self-testing means that $p(a,b|x,y)$ guarantees the existence of a $\Phi$ such that
\begin{equation}
\begin{split}
&\Phi{|\psi\rangle}_{AB}{|00\rangle}_{A'B'}={|junk\rangle}_{AB}{|\Phi^{+}\rangle}_{A'B'}\,,\\
&\Phi A_xB_y{|\psi\rangle}_{AB}{|00\rangle}_{A'B'}={|junk\rangle}_{AB}{(\vec a_x\cdot\vec\sigma\otimes\vec b_y\cdot\vec\sigma)|\Phi^{+}\rangle}_{A'B'}\,.
\end{split}
\end{equation}
It is worthwhile to note the isometry here indeed is a virtual protocol \cite{refth,refjd,refxy}. It takes care of the degeneracy of the problem with respect to local unitaries and the addition of irrelevant degrees of freedom. All
that need to be done in the laboratory is to query the boxes and derive $p(a,b|x,y)$.

Similarly, as we wrote, all we can say \textit{a priori} about the measurements is that they are defined by some otherwise unknown projectors. At the mathematical level, we are going to work with the derived observables $A_x=\Pi_{a=+1}^x-\Pi_{a=-1}^x$, $B_x=\Pi_{b=+1}^y-\Pi_{b=-1}^y$. These are just convenient ways of doing algebra with the projectors, and should not be seen as assumptions about the actual implementation of the devices (for instance, $A_x^2$ won't refer to the possibility of performing the measurement twice sequentially).

\section{Self-testing of the singlet: known results} 

\subsection{The isometry and its control operators}

\begin{figure}[htb!]
 \centering
 \includegraphics{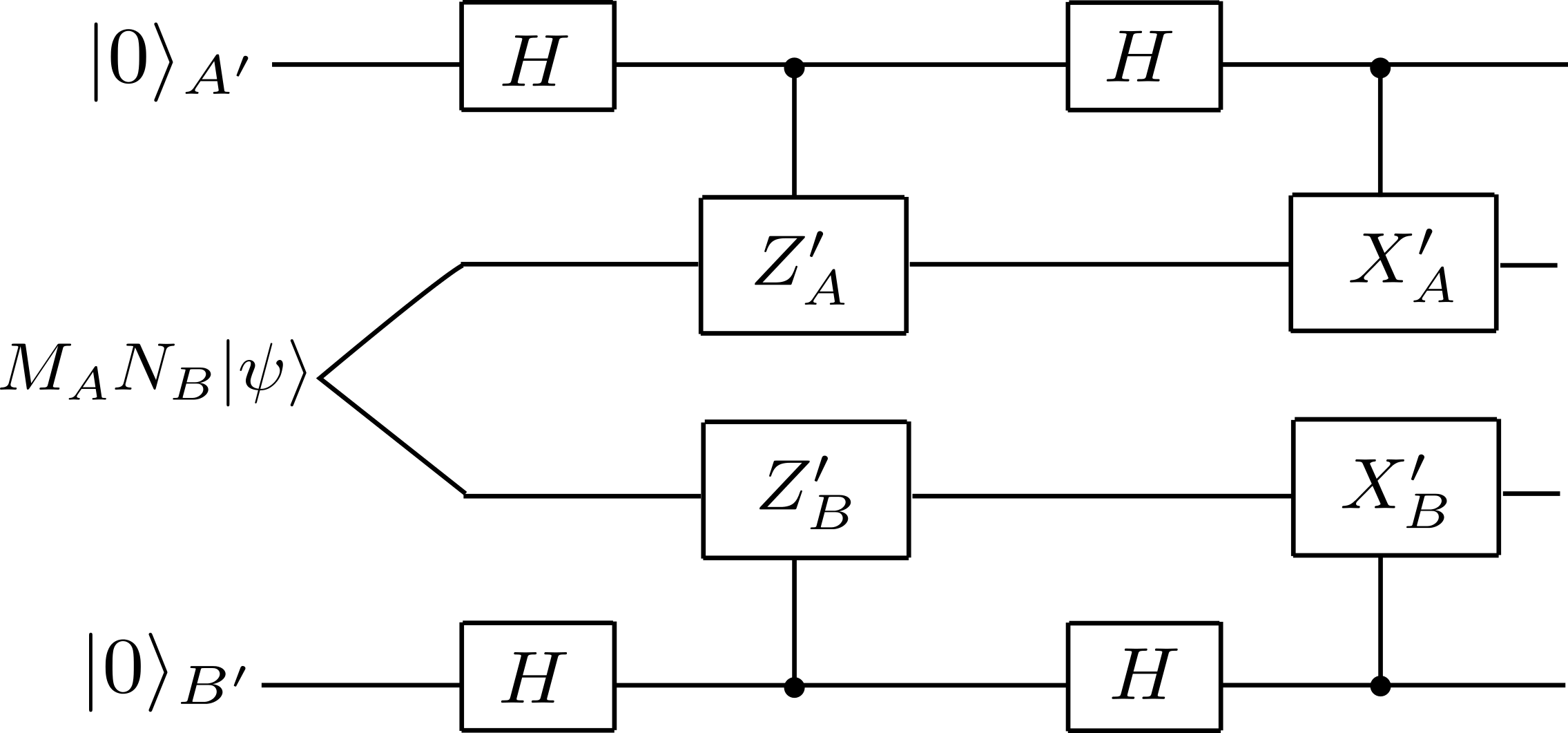}
 \caption{\label{fig1}Local isometry $\Phi$ that allows self-testing of the singlet and the local operators.  $H$ are Hadamard gates while  $Z'_{A/B}$ and $X'_{A/B}$ are the control $Z_{A/B}$ and control $X_{A/B}$ gates up to a local unitary, respectively. These control gates put the
relevant degrees of freedom from the self-tested system into trusted auxiliary qubits $|00\rangle_{A'B'}$,  which has the right dimension for the desired state and measurement to be possible.}
\end{figure}

There is no reason why the construction of the isometry should be unique, but as it turns out, all the works reporting self-testing of the singlet so far have used the same construction, depicted in Fig.~\ref{fig1}. The gates coupling the unknown system with the ancilla are controlled on the side of $A$ and $B$ by the \textit{control operators} $X'$ and $Z'$, which are themselves unknown. It has been shown \cite{refmm,slovaca} that the singlet is self-tested if these operators are unitary and satisfy
\begin{equation}\label{nc}
\begin{split}
& Z'_A|\psi\rangle=Z'_B|\psi\rangle\,,\\
& X'_A|\psi\rangle=X'_B|\psi\rangle\,,\\
&X'_AZ'_A|\psi\rangle=-Z'_AX'_A|\psi\rangle\,,\\
&X'_BZ'_B|\psi\rangle=-Z'_BX'_B|\psi\rangle\,.\\
\end{split}
\end{equation}   
A proof of self-testing of the singlet may then consist of constructing the control operators $X'_{A/B}$ and $Z'_{A/B}$ from $\Pi_{a}^x,\Pi_{b}^y$, then use the knowledge of $p(a,b|x,y)$ to show that those operators are indeed unitary and satisfy \eqref{nc}. 

We are going to review quickly the two known criteria for self-testing in this perspective (see \cite{refmm,slovaca} for the detailed proof of all the claims).

\subsection{CHSH criterion}

The CHSH criterion for self-testing of the singlet is:

\emph {Consider the (2,2,2) scenario with unknown operators $\{A_0,A_1;B_0,B_1\}$, if a CHSH test yields $S= \langle A_0B_0\rangle+\langle A_0B_1\rangle+\langle A_1B_0\rangle -\langle A_1B_1\rangle=2\sqrt{2}$ exactly, then up to a local isometry, the state is a singlet and the measurements are the corresponding Pauli operators.}

In a sense, self-testing can be achieved from a single number, $S=2\sqrt{2}$: the only quantum point that achieves this violation is such that $\langle A_0B_0\rangle=\langle A_1B_0\rangle=1/\sqrt{2}$ and $\langle A_0B_1\rangle=-\langle A_1B_1\rangle=1/\sqrt{2}$. With these relations, one can prove that the control operators $Z'_A=A_0$, $X'_A=A_1$, $Z'_B=\frac{B_0-B_1}{\sqrt{2}}$ and $X'_B=\frac{B_1+B_0}{\sqrt{2}}$ are indeed unitary and satisfy \eqref{nc}.

\subsection{Mayers-Yao criterion}

The original Mayers-Yao criterion was phrased in the (2,3,2) scenario, but it was noticed later \cite{refmm} that the third measurement is needed only on one side, say Bob's.

\emph{Consider five unknown operators $\{A_0,A_1;B_0,B_1,B_2\}$ with binary outcomes and $[A_x,B_y]=0$. If the following correlations are observed
\begin{equation}
\begin{split}
&\langle \psi| A_0B_0|\psi\rangle=\langle \psi| A_1B_1|\psi\rangle=1\\
&\langle \psi| A_0B_1|\psi\rangle=\langle \psi| A_1B_0|\psi\rangle=0\\
&\langle \psi| A_0B_2|\psi\rangle=\langle \psi| A_1B_2|\psi\rangle=1/\sqrt{2}\\
\end{split}
\end{equation}
then up to a local isometry, the state is a singlet and the measurements are the suitable complementary Pauli operators.}
 
The first two conditions  can be rewritten as $A_0|\psi\rangle=B_0|\psi\rangle$ and $A_1|\psi\rangle=B_1|\psi\rangle$; the third setting $B_2$ is needed to derive the anti-commutation relations  $A_0A_1|\psi\rangle=-A_1A_0|\psi\rangle$, $B_0B_1|\psi\rangle=-B_1B_0|\psi\rangle$. Then $A_0,A_1,B_0,B_1$ directly define the suitable control operators to self-test the singlet.

\section{Main result}

Our main result is the following theorem, which characterizes all the criteria that self-test the singlet and the corresponding measurements for the (2,2,2) scenario.

\newtheorem*{lem}{Theorem 1}
\begin{lem}
 Consider four unknown operators $\{A_0,A_1;B_0,B_1\}$ with binary outcomes labeled $\pm 1$ and assumed to fulfill $[M_A,N_B]=0$. The observed correlations $E_{xy}\equiv \langle \psi| A_xB_y|\psi\rangle$ self-test the singlet if and only if they satisfy one of the eight conditions
\begin{equation}\label{t5}
\sum_{(x,y)\neq (i,j)}\arcsin(E_{xy})-\arcsin(E_{ij})=\xi\pi \quad \textrm{with $i,j\in\{0,1\}$, $\xi\in\{+1,-1\}$}
\end{equation}
with $\arcsin(E_{xy})_{x,y\in\{0,1\}}\in [-\frac{\pi}{2},\frac{\pi}{2}]$, and provided $\arccos(E_{xy})=0\text{ or }\pi$ holds for at most one pair $(x,y)$. 
\end{lem}

\begin{proof} The conditions \eqref{t5} are known to define the boundary of the correlations $\{E_{xy}\}$ achievable with quantum physics in the (2,2,2) scenario \cite{refcs,refld}. More importantly for us, it follows from the work of Masanes \cite{web1,web2} that an extremal point of the (2,2,2) quantum set can be generated by measuring a singlet if and only if it satisfies \eqref{t5} (see \ref{apponlyif} for the spelled-out argument). It is easily checked that all those points are nonlocal if and only if at least three of the $\arccos(E_{xy})$ are not zero or $\pi$(see \ref{appalpha}). Since ideal self-testing can only be done with extremal points of the quantum set, this settles the ``only if" part of the proof: these correlations are the only candidates to be self-test criteria for the singlet. Now we need to prove that all of them actually are, i.e., that all those points can be achieved only by measuring the singlet. We'll do it by constructing explicit control operators from $\{A_0,A_1;B_0,B_1\}$.

The starting point is a geometric observation. Since the operators $A_x$ and $B_y$ are unitary on $\ket{\psi}$, the four vectors $\vec{A}_x=A_x|\psi\rangle$ and $\vec{B}_y=B_y|\psi\rangle$ are unit vectors defined in some unknown dimension. Let us denote $E_{xy}\equiv {\vec{A}_x}^\dagger \vec{B}_y =\cos\alpha_{xy}$ with $\alpha_{xy}\geq 0$. Given their scalar products with $\vec{B}_0$, the angle $\theta$ between $\vec{A}_0$ and $\vec{A}_1$ must satisfy
\begin{equation}
\label{boundtheta1}
|\alpha_{00}-\alpha_{10}| \leq \theta \leq \alpha_{00}+ \alpha_{10}\,.
\end{equation} The lower and the upper bound are reached when $\vec{B}_0$ lies in the same plane (on either side, or between the two). Similarly, because of the scalar products with $\vec{B}_1$, one has 
\begin{equation}
\label{boundtheta2}
|\alpha_{01}-\alpha_{11}| \leq\theta\leq \alpha_{01}+\alpha_{11}\,.
\end{equation}

Now, the eight equalities in \eqref{t5} are equivalent in the sense that each one can be transformed into the other by relabelling measurements and/or outcomes \cite{web1}. So, without loss of generality we consider $i=0$, $j=1$ and $\xi=+1$. The equality can be rewritten as $\arccos(E_{00})+\arccos(E_{10})=\arccos(E_{01})-\arccos(E_{11})$, that is 
\begin{equation}\alpha_{00}+ \alpha_{10}=\alpha_{01}-\alpha_{11}\,.\label{alphas}\end{equation} Together with \eqref{boundtheta1} and \eqref{boundtheta2}, this implies that $\theta=\alpha_{00}+ \alpha_{10}=\alpha_{01}-\alpha_{11}$ and in particular that all the four vectors lie in the same plane in the positions sketched in Fig.~(\ref{fig2}). In particular we have
\begin{align}
\label{r1}
B_0|\psi\rangle = \frac{\sin(\alpha_{00})A_1|\psi\rangle+\sin(\alpha_{10})A_0|\psi\rangle}{\sin(\alpha_{00}+\alpha_{10})},\\
\label{r2}
A_1|\psi\rangle = \frac{\sin(\alpha_{10})B_1|\psi\rangle+\sin(\alpha_{11})B_0|\psi\rangle}{\sin(\alpha_{11}+\alpha_{10})}.
\end{align}

\begin{figure}[htb!]
 \centering
 \includegraphics{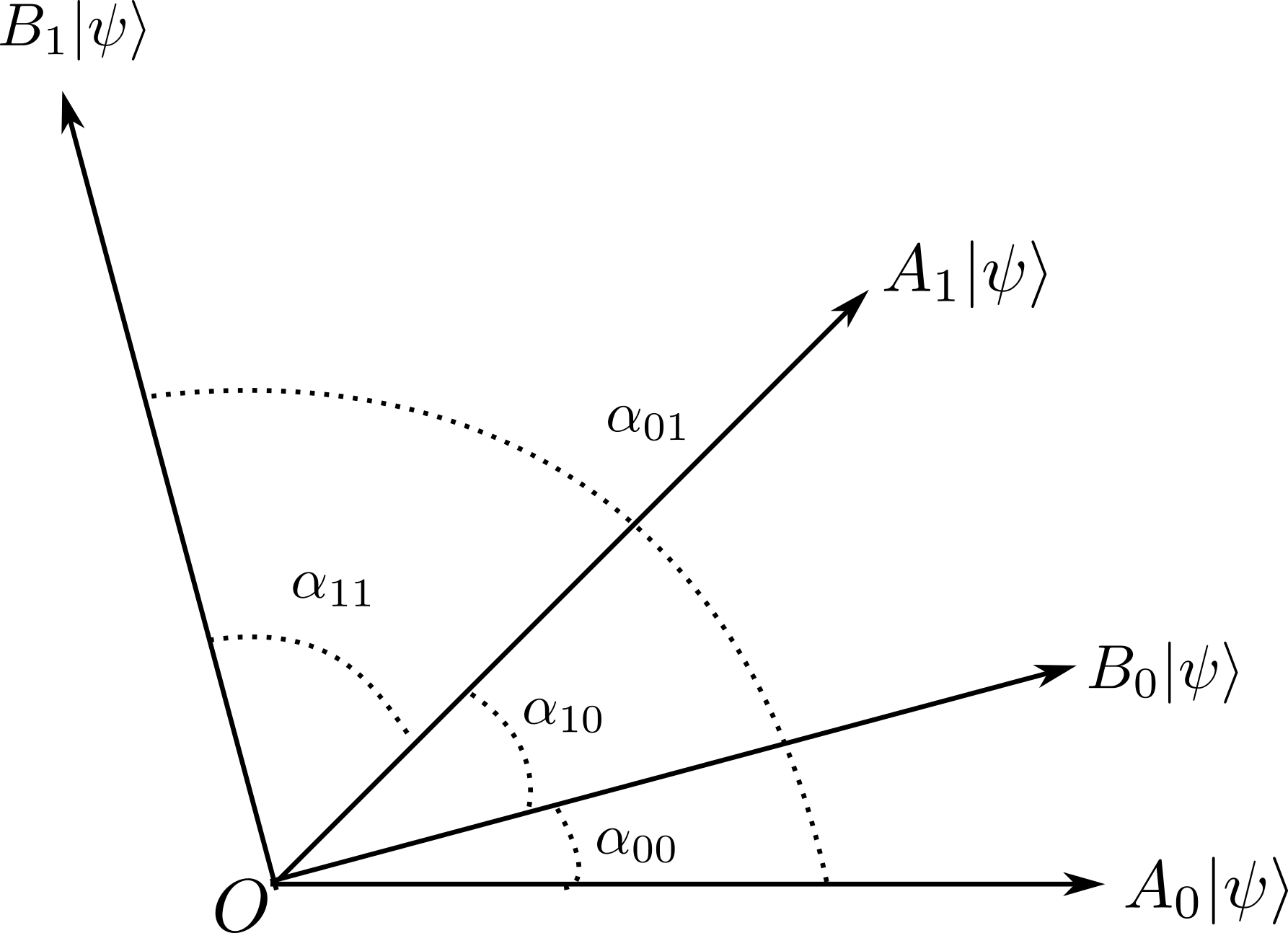}
 \caption{\label{fig2}$A_0|\psi \rangle$, $A_1| \psi \rangle$, 
$B_0|\psi \rangle$ and $B_1| \psi \rangle$ in the same plane when $\alpha_{00}+\alpha_{10}=\alpha_{01}-\alpha_{11}$. }
 \end{figure}

Now, using $[M_A,N_B]=0$ we obtain\\
\begin{equation}
\begin{split}
 &B^{2}_0|\psi\rangle=B_0\frac{\sin(\alpha_{00})A_1|\psi\rangle+\sin(\alpha_{10})A_0|\psi\rangle}{\sin(\alpha_{00}+\alpha_{10})}\\
 &=\frac{\sin(\alpha_{00})A_1+\sin(\alpha_{10})A_0}{\sin(\alpha_{00}+\alpha_{10}))}B_0|\psi\rangle\,=\,\left(\frac{\sin(\alpha_{00})A_1+\sin(\alpha_{10})A_0}{\sin(\alpha_{00}+\alpha_{10})}\right)^{2}|\psi\rangle\,.
\end{split}
\end{equation}
Since $B^{2}_0=A^{2}_0=A^{2}_1=I$, we obtain the anti-commutation relations
 \ba\label{antic1}
 (A_1A_0+A_0A_1)|\psi\rangle=2\cos(\alpha_{00}+\alpha_{10})|\psi\rangle
 \ea provided $\alpha_{00}\neq 0$ and $\alpha_{10}\neq 0$. Similarly, starting from $A_1^2|\psi\rangle$, we'd obtain
\ba\label{antic2}
(B_1B_0+B_0B_1)|\psi\rangle=2\cos(\alpha_{11}+\alpha_{10})|\psi\rangle
\ea
In \ref{appalpha} we show that these anti-commutation relations can be obtained as long as at most one of the four $\alpha_{xy}$ is zero.

The anti-commutation relations are the key element to conclude the proof. They enable to show that the control operators
\begin{equation}\label{A12}
\begin{split}
 &Z'_A=A_0\\
 &X'_A=\frac{A_1-\cos(\alpha_{00}+\alpha_{10})A_0}{\sin(\alpha_{00}+\alpha_{10})}\\
 &Z'_B=\frac{\sin(\alpha_{01})B_0-\sin(\alpha_{00})B_1}{\sin(\alpha_{01}-\alpha_{00})}\\
 &X'_B=\frac{\cos(\alpha_{00})B_1-\cos(\alpha_{01})B_0}{\sin(\alpha_{01}-\alpha_{00})}\\
\end{split}
\end{equation}
satisfy the conditions \eqref{nc}; self-testing then follows as in \cite{refmm}.

Indeed, since  $X'_AZ'_A|\psi\rangle+Z'_AX'_A|\psi\rangle
   =(A_1A_0+A_0A_1-2\cos(\alpha_{00}+\alpha_{10}))/ \sin(\alpha_{00}+\alpha_{10}))|\psi\rangle$, and
 $X'_BZ'_B|\psi\rangle+Z'_BX'_B|\psi\rangle
   =\sin(\alpha_{01}+\alpha_{00})(B_1B_0+B_0B_1-2\cos(\alpha_{01}-\alpha_{00}))/ \sin^2(\alpha_{01}-\alpha_{00}))|\psi\rangle$, it follows from
\eqref{antic1} and \eqref{antic2} that
 \begin{equation}
 \begin{split}\label{A8}
 &X'_AZ'_A|\psi\rangle=-Z'_AX'_A|\psi\rangle\\
&X'_BZ'_B|\psi\rangle=-Z'_BX'_B|\psi\rangle\\
\end{split}
\end{equation}
Besides, it holds $\langle \psi| Z'_AZ'_B|\psi\rangle=\langle \psi| X'_AX'_B|\psi\rangle=1$, which implies  
\begin{equation}\label{A9}
\begin{split}
&Z'_A|\psi\rangle=Z'_B|\psi\rangle \\
&X'_A|\psi\rangle=X'_B|\psi\rangle\\
\end{split}
\end{equation}
provided we can prove that the control operators are unitary when acting on $|\psi\rangle$. For $Z'_A$, it is the case by definition.
For $X'_A$, using relations \eqref{antic1} we find
\begin{equation} \langle \psi|(X'_A)^{+}X'_A|\psi\rangle = \langle \psi|\frac{A^2_1+\cos^2(\alpha_{00}+\alpha_{10})A^2_0-\cos(\alpha_{00}+\alpha_{10})(A_0A_1+A_1A_0)}{\sin^2(\alpha_{00}+\alpha_{10})}|\psi\rangle = 1\,,\end{equation} and similarly for  $Z'_B|\psi\rangle$ and $X'_B|\psi\rangle$. \end{proof}

\begin{figure}[htb!]
 \centering
 \includegraphics{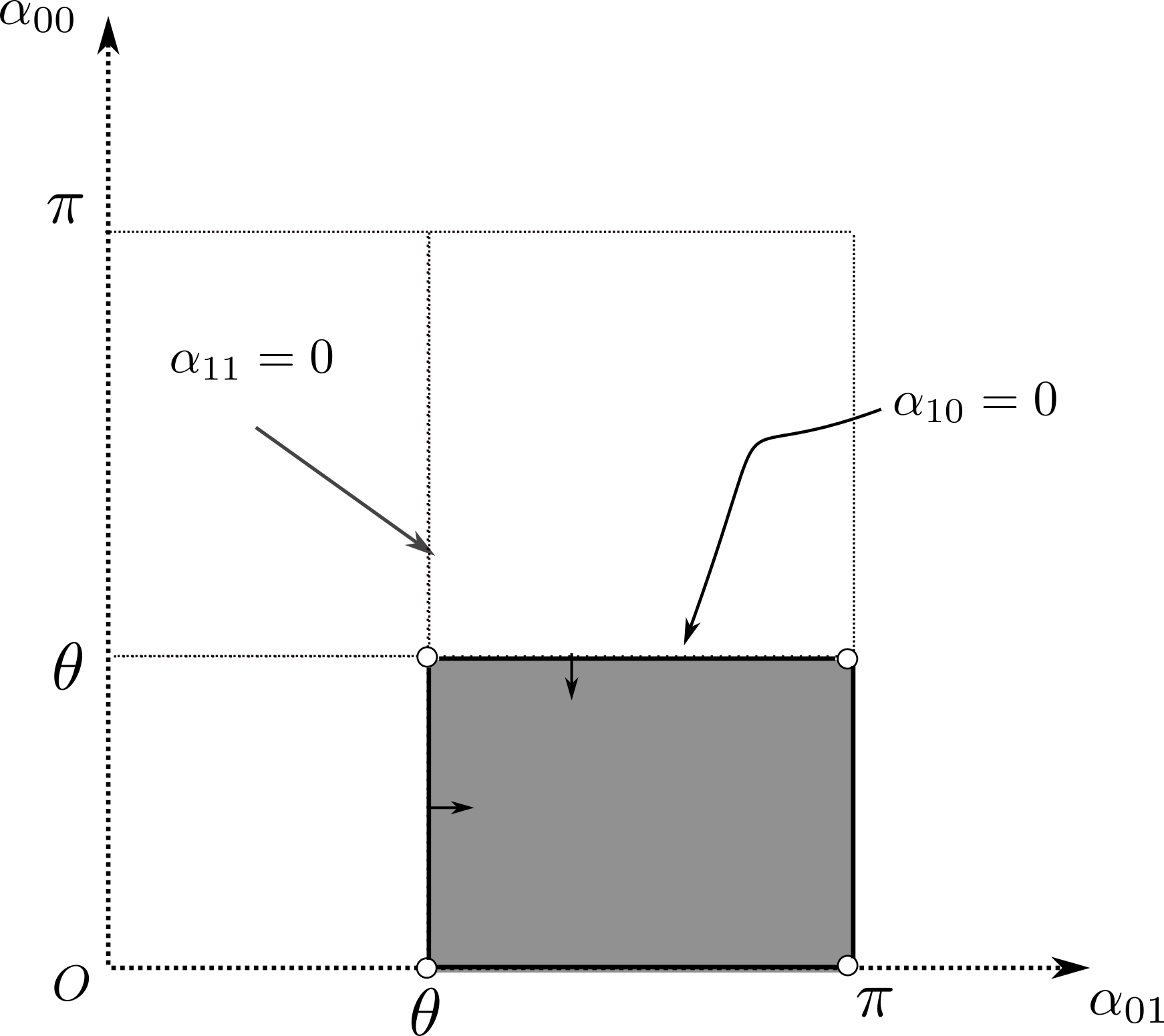}
 \caption{\label{fig3} Feasible points for self-testing singlet for a given $\theta$. $\theta$ takes the value in interval $[0,\pi]$. The feasible domain is the colored square, excluding the four vertices.}
\end{figure}

\begin{figure}[htb!]
 \centering
 \includegraphics{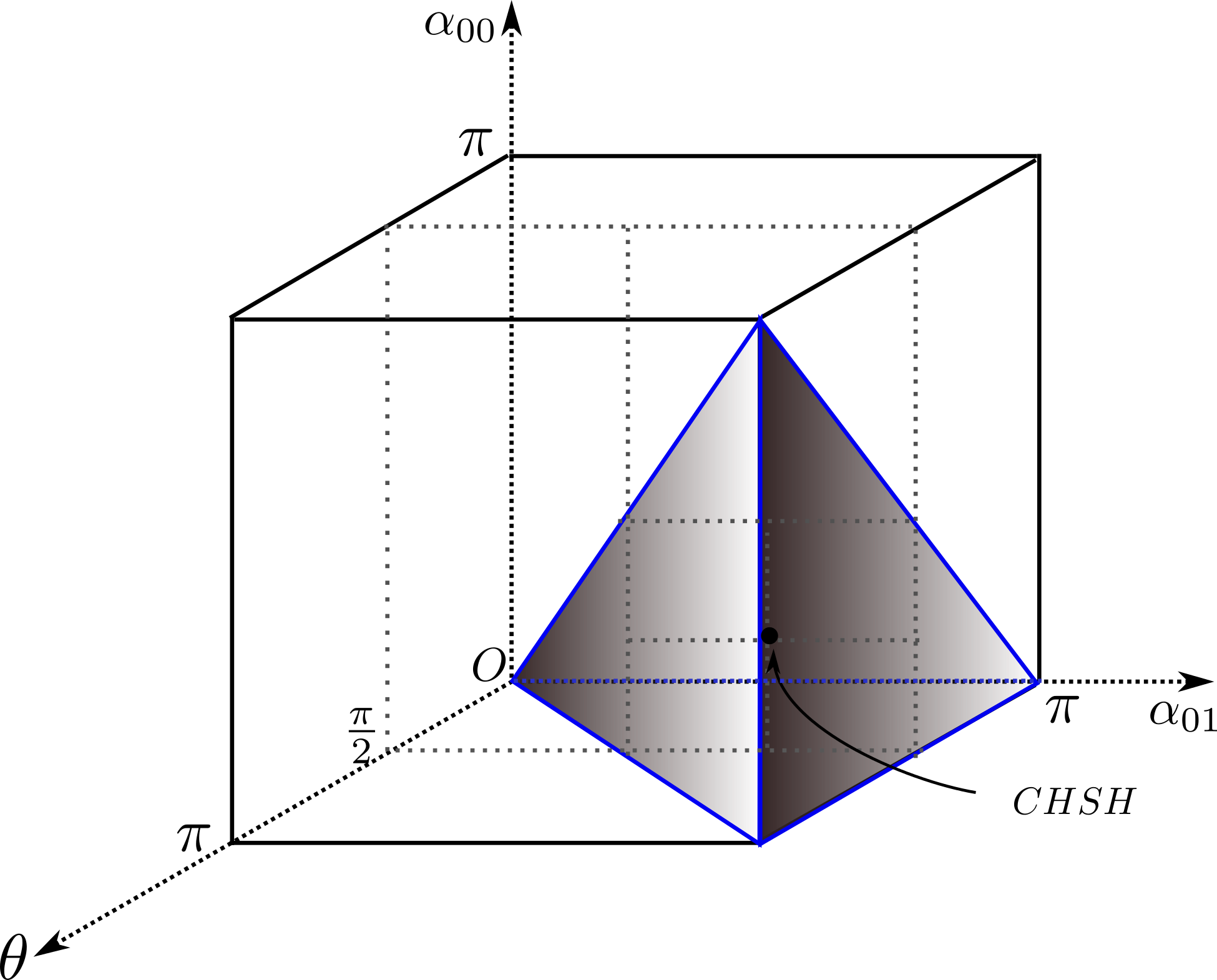}
 \caption{\label{fig4}  All the feasible points that can be used for self-testing singlet and the corresponding local operators, for condition $\alpha_{00}+\alpha_{10}=\alpha_{01}-\alpha_{11}$.}
\end{figure}

As we noted in the proof, the eight equalities in \eqref{t5} are equivalent under suitable relabelling. So we can focus on the equality \eqref{alphas} to show the region of parameters in which the singlet can be self-tested. Again, denote $\theta=\alpha_{00}+\alpha_{10}=\alpha_{01}-\alpha_{11}$. We have $\theta$ belongs to $[0,\pi]$ by definition. For fixed $\theta$, self-testing defines a rectangle in the $(\alpha_{00},\alpha_{01})$ plane, excluding the vertices because there two of the $\alpha$'s are zero (Fig.~\ref{fig3}). By plotting $\theta$ on a third axis, the set of points that self-test the singlet forms a tetrahedron without its edges (Fig.~\ref{fig4}).


The CHSH criterion is the point $\alpha_{00}=\alpha_{10}=\alpha_{01}=-\alpha_{11}=\frac{\pi}{4}$ on the plane $\theta=\frac{\pi}{2}$. The Mayers-Yao criterion can be seen as taking the vertex $\alpha_{00}=\alpha_{11}=0$, $\alpha_{10}=\alpha_{01}=\frac{\pi}{2}$ on the same plane, which does not self-test, and adding a third setting in same plane on (say) Bob's side satisfying $\alpha_{02}=\alpha_{12}=\frac{\pi}{4}$. For perfect self-testing, this is now clearly seen to be redundant: one could just replace $B_0$ with $B_2$ and obtain the point $\alpha_{00}=\alpha_{10}=\frac{\pi}{4}$, $\alpha_{01}=\frac{\pi}{2}$ and $\alpha_{11}=0$ which is again on the $\theta=\frac{\pi}{2}$ plane and does self-test.

\section{Robustness of the self-testing criteria}

Ideal self-testing won't ever be possible because of experimental imperfections. Thus, it is important to discuss how the self-test criteria perform in case of deviations
\begin{equation}
\begin{split}\label{eq:cd}
&|\langle \psi| A_xB_y|\psi\rangle-E^{\textrm{ideal}}_{xy}|\leq \varepsilon\\
\end{split}
\end{equation}
from the ideal values. We base this robustness study on the techniques introduced in \cite{refth}.

\subsection{The swap technique}

The isometry in Fig.~\ref{fig1} can be re-written as a \textit{swap} operator $S_{AA}\otimes S_{BB'}$ with
$S_{AA'}=U_{AA'}V_{AA'}U_{AA'}$ and
\begin{equation}\label{defswap}
\begin{split}
&U_{AA'}=I_{A}\otimes |0\rangle \langle 0|+X'_{A}\otimes|1\rangle \langle 1| \\
&V_{AA'}=\frac{I_{A}+Z'_{A}}{2}\otimes I+\frac{I_{A}-Z'_{A}}{2}\otimes \sigma_{x}\\
\end{split}
\end{equation} and the same for $S_{BB'}$  \cite{refMI}. If the ancilla qubits are prepared in the state $\ket{0}$, the first application of $U_{AA'}$ is equivalent to the identity. After this isometry, the trusted auxiliary systems will be left in the state $\rho_{swap}=tr_{AB}[S\rho_{AB}\otimes |00\rangle \langle 00|_{A'B'}S^{+}]=\sum_{ikjl}C_{ikjl}|i\rangle \langle k|\otimes |j\rangle \langle l|$ where $C_{ikjl}=\frac {1}{16}tr_{AB}[(I+Z'_A)^{1-k}(X'_A-Z'_AX'_A)^{k}(I+Z'_A)^{1-i}(X'_A-X'_AZ'_A)^{i}\otimes (I+Z'_B)^{1-l}(X'_B-Z'_BX'_B)^{l}(I+Z'_B)^{1-j}(X'_B-X'_BZ'_B)^{j}\rho_{AB}]$ for $i,j,k,l \in\{0,1\}$. Then one can describe how close $\rho_{swap}$ is to $\ket{\Phi^+}$ using the fidelity
\begin{equation}
\begin{split}
F&=\langle\Phi^{+}|\rho_{swap}|\Phi^{+}\rangle\,.\\
\end{split}
\end{equation} All that needs to be done now is to find a lower bound of $F$. If one knows how to relate the control operators to the actual operators $A_x,B_y$, some terms in $F$ are determined by the observed correlations \eqref{eq:cd}. The terms that are not determined should be compatible with a quantum realisation. As well known, this last requirement can't be formulated as an efficient constraint, but it can be relaxed to a family of semi-definite constraints $\Gamma_{s,s'}$ \cite{refMS,refMN}. Since the objective function $F$ is also linear, the optimisation can then be cast as a semi-definite program (SDP). In general, one needs an infinite family to recover the exact quantum bound; by choosing a finite family, one is optimising over a set that is larger than the quantum set.

The control operators defined above \eqref{A12} seem to be a good guess. However, direct replacement is not possible: as soon as $\varepsilon>0$, the anti-commutation relations \eqref{antic1} and \eqref{antic2} won't be satisfied exactly. As a consequence, in general the operators defined on the r.h.s. of \eqref{A12} can't be guaranteed to be unitary. More specifically, $Z'_A=A_0$ remains unitary. By redefining the target state with a suitable rotation on Bob's system, $\ket{\Phi^+}\longrightarrow \ket{\Phi}=\cos(\frac{\alpha_{00}}{2})\frac{|00\rangle+|11\rangle}{\sqrt{2}}+\sin(\frac{\alpha_{00}}{2})\frac{|01\rangle-|10\rangle}{\sqrt{2}}$, one can set $Z'_B=B_0$ and $X'_B=\frac{B_1-\cos(\alpha_{10}+\alpha_{11})B_0}{\sin(\alpha_{10}+\alpha_{11})}$ in \eqref{A12}. So $Z'_{A/B}$ can always be defined as to be unitary, but we have to deal with $X'_A$ and $X'_B$ being possibly not unitary. One way around this obstacle is to use the method of ``localizing matrices'', already used in the context of self-testing \cite{refjd}. The idea is to invoke two new operators $A_2$ and $B_2$ satisfying $A^2_2 =B^2_{2}=I$, and simply set $X_A'=A_2$ and $X_B'=B_2$. Then one has to relate these control operators to the actual ones, capturing the idea that the new definition can't be too different from that of the ideal case \eqref{A12}. One way to do this, not guaranteed to be optimal, consists in imposing
\begin{equation}\label{auxi}
\begin{split}
&A_2\,[A_1-\cos(\alpha_{00}+\alpha_{01})A_0]\equiv A_2\tilde{X}_A\geq 0\\
&B_2\,[B_1-\cos(\alpha_{10}+\alpha_{11})B_0]\equiv B_2\tilde{X}_B\geq 0\\
\end{split}
\end{equation} which are enforced by imposing that the localizing matrices $\Gamma(A_2\tilde{X}_A)_{j,j'}$ and $\Gamma(B_2\tilde{X}_B)_{j,j'}$ are Hermitian and positive semi-definite. We use the simplest non-trivial form of the localizing matrix, which is the $4\times 4$ matrix $\Gamma(A_2\tilde{X}_A)_{j,j'}=\langle{\cal A}_j (A_2\tilde{X}_A) {\cal A}_{j'}\rangle$ with ${\cal A}=(I,A_0,A_1,A_2)$; and similarly for Bob.



\subsection{Robustness bounds for various criteria}

Now we have collected all the general tools needed to present some robustness bounds. For simplicity, we focus on a single parameter family $(\theta=\pi/2,\alpha_{00}=\pi/4,\alpha_{01})$. For $\alpha_{01}=3\pi/4$, this is the CHSH criterion; for $\alpha_{01}=\pi/2$, this can be seen as a four-setting version of the Mayers-Yao test ($A_1=B_1$, $A_0$ orthogonal to $A_1$ and $B_0$ between them at forty-five degrees). For these parameters, \eqref{A12} yields $X'_A=A_1$, so we can avoid introducing $A_2$; for CHSH, $X'_B=B_1$ also holds, whereas for the other cases we have to use $X'_B=B_2$ and add the corresponding localising matrix. For comparison, we also consider the Mayers-Yao criterion with three settings on Bob's side (which, like CHSH, does not need a localizing matrix). 

\begin{figure}[htb!]
 \centering
 \includegraphics{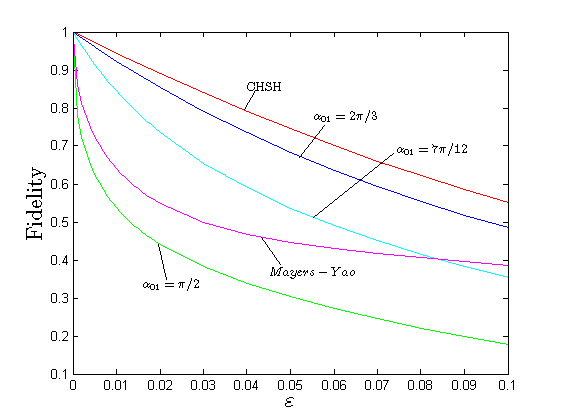}
 \caption{\label{fidelity} Lower bound for the certifiable singlet fidelity $F$ as a function of the imprefections of the observed correlations $\varepsilon$ \eqref{eq:cd}. We plot the bounds for four four-setting criteria $(\theta=\pi/2,\alpha_{00}=\pi/4,\alpha_{01})$ and for the five-setting Mayers-Yao criterion.}
\end{figure}

We run the SDP optimisation including all the first momenta $A_x,B_y$, all the second momenta $A_xA_{x'},B_yB_{y'},A_xB_{y}$; for simplicity of the code, we include only the third momenta $A_0A_1A_0$ and $B_0B_{1/2}B_0$ that appear in the expression of the fidelity. The result is presented in Fig.~\ref{fidelity} \footnote{Notice that the CHSH and Mayers-Yao bounds are \textit{a priori} different from those plotted in Fig.~3 of \cite{refjd}. There, the constraints of the SDP were all the statistics obtained by measuring a Werner, while here we use the more relaxed conditions \eqref{eq:cd} that do not specify the marginals.}.

We note that the curves, while valid lower bounds, are not guaranteed to be tight. First, as we mentioned, the isometry may not be optimal, and in fact for CHSH we know that one can do better (Fig.~2 of \cite{refjd}). Then, we worked at a finite level of the SDP hierarchy. Finally, when used, the localizing matrix method may add to the lack of tightness. The resulting approximation may even vary from one criterion to another. Thus, while it seems that the CHSH criterion remains the most robust, comparisons should not be considered as conclusive.

\section{Conclusions}

We have studied the problem of self-testing of the two-qubit singlet and the corresponding local operators in a device-independent scenario. Two inequivalent criteria were known previous to this work. Here we prove that, in the case of two input per party and two output per measurement [(2,2,2) scenario], all the extremal quantum points achievable with the singlet also self-test it (i.e., those points can be achieved only with projective measurements on the singlet).

The work of Miller and Shi sheds light on our result from a different perspective \cite{millershi}. They provided necessary and sufficient conditions for a \textit{binary XOR game} to be self-testing. The criteria we discovered here use only correlation functions $E_{xy}$ and can thus be cast into such a game --- although, at this stage, this would be a useless processing of information \cite{pironio,samedata}. From this point of view, our contribution consists in listing explicitly all the binary XOR games that self-test the singlet (see \ref{appgames}) and proving that only such games do the job, in the (2,2,2) scenario. A thorough study of these questions remains open for other states and scenarios.

\section*{Acknowledgements } We acknowledge useful suggestions from Jean-Daniel Bancal, Cai Yu and Matthew McKague. This work is funded by the Singapore Ministry of Education (partly through the Academic Research Fund Tier 3 MOE2012-T3-1-009), by the National Research Foundation of Singapore, by the National Natural Science Foundation of China (Grants No. 61272057, No. 61170270) and by the Beijing Higher Education Young Elite Teacher Project (Grants No. YETP0475 and No. YETP0477). Yukun Wang is also supported in part by the Government of China through China Scholarship Council.

\appendix

\section{Extremal points of the (2,2,2) quantum set achievable by measuring the singlet}\label{apponlyif}

We want to prove that the only extremal points of the (2,2,2) quantum set that can be obtained by measuring a singlet are those that satisfy \eqref{t5}.

First, we notice that all the POVMs with binary outcomes are commuting. The statistics of the POVM can be obtained by performing a projective measurement, then process the outcome with classical noise (see Lemma 1 in Ref.~\cite{web2}). Thus, we can restrict to projective measurements to find the extremal points.

Now, for projective local measurements on the singlet, it holds $\langle A_x\rangle = \langle B_y\rangle=0$: thus, the only extremal points that the singlet can generate must have unbiased marginals and involve only the correlators $\langle A_xB_y\rangle$. The necessary and sufficient condition that correlators must satisfy to define an extremal point is precisely \eqref{t5}, and all these points can be in fact obtained by measuring the singlet (see Lemma 4 in Ref.~\cite{web1} and its proof).

\section{Cases when some of the $\alpha_{xy}$ are equal to zero}\label{appalpha}

We work under condition \eqref{alphas}. We have claimed that the anti-commutation relations \eqref{antic1} and \eqref{antic2}, fundamental to prove self-testing, can be obtained if one of the $\alpha_{xy}$ is zero and can't be obtained if two are (or all four, of course). In this appendix, we study thoroughly these cases.

\begin{itemize}
\item
If (say) $\alpha_{00}=0$, then \eqref{antic1} cannot be obtained from $B_0^2|\psi\rangle$ as we did in the main text because the term containing the anti-commutator would be multiplied by zero. However, from $B_1^2|\psi\rangle$ it would follow that
\ba
(A_1A_0+A_0A_1)|\psi\rangle=2\cos(\alpha_{01}-\alpha_{11})|\psi\rangle\label{A1}
\ea which is exactly \eqref{antic1} because of \eqref{alphas}. The other cases are similar. So, if only one of the $\alpha_{xy}$ is zero, we can finish the proof of self-testing.

For consistency, let's check that the correlations violate CHSH in this case. Our version of CHSH reads $C=\cos\alpha_{00}-\cos\alpha_{01}+\cos\alpha_{10}+\cos\alpha_{11}$. If $\alpha_{00}=0$, we have $\alpha_{10}=\alpha_{01}-\alpha_{11}$. A violation of CHSH is then equivalent to $\cos(\alpha_{01}-\alpha_{11})+\cos\alpha_{11}-\cos\alpha_{01}>1$, which is readily verified to be true as long as $\alpha_{01}$ and $\alpha_{11}$ are neither $0$ or $\pi$. The strict proof is also shown below,
\begin{align*}
    &-\cos\alpha_{01}+\cos\alpha_{10}+\cos\alpha_{11}\\
    &=1-2\cos^2\frac{\alpha_{01}}{2}+2\cos\frac{\alpha_{10}+\alpha_{11}}{2}\cos\frac{\alpha_{10}-\alpha_{11}}{2}\\
    &=1-2\cos^2\frac{\alpha_{01}}{2}+2\cos\frac{\alpha_{01}}{2}\cos\frac{\alpha_{10}-\alpha_{11}}{2}\\
    &>1-2\cos^2\frac{\alpha_{01}}{2}+2\cos\frac{\alpha_{01}}{2}\cos\frac{\alpha_{01}}{2}\\
    &=1,
\end{align*}
where the greater sign is due to the fact that $-\frac{\alpha_{01}}{2}<\alpha_{10}-\frac{\alpha_{01}}{2}=\frac{\alpha_{10}-\alpha_{11}}{2}=\frac{\alpha_{01}}{2}-\alpha_{11}<\frac{\alpha_{01}}{2}$, and the property of the $\cos$ function in $[-\pi,\pi]$.

Similarly, the violation of CHSH in the case of $\alpha_{01}=\pi$ can also be certified.

\item If two of the $\alpha_{xy}$ are zero, we don't expect self-testing to be possible since the observed correlations would be local. Indeed:
\begin{enumerate}
\item $\alpha_{00}=\alpha_{10}=0\;\Longrightarrow\; A_0=B_0=A_1$ (here and below, acting on $\ket{\psi}$ is implicit).\\
\item $\alpha_{00}=\alpha_{01}=0\;\Longrightarrow\; A_0=B_0=B_1$.\\
\item $\alpha_{00}=\alpha_{11}=0\;\Longrightarrow\; A_0=B_0\,,\,A_1=B_1$.\\
\item $\alpha_{01}=\alpha_{10}=0\;\Longrightarrow\; A_0=B_1\,,\,A_1=B_0$.\\
\item $\alpha_{01}=\alpha_{11}=0\;\Longrightarrow\; A_0=B_1=A_1$.\\
\item $\alpha_{10}=\alpha_{11}=0\;\Longrightarrow\; A_1=B_0=B_1$.\\
\item $\alpha_{00}=\alpha_{01}=\alpha_{10}=\alpha_{11}=0\;\Longrightarrow\; A_0=A_1=B_0=B_1$.
\end{enumerate}
Consider for instance case (ii): because $\alpha_{00}=0$, \eqref{antic1} cannot be obtained from $B_0^2|\psi\rangle$; and because $\alpha_{01}=0$, the equivalent \eqref{A1} cannot be obtained from $B_1^2|\psi\rangle$. In the same way, (vi) prevents to obtain the  relation for A, (i) and (v) that for B, and (iii) and (iv) both. For (vii), though the  relation for A, B can be get, however we still couldn't define the suitable control operators.
\end{itemize}

\section{Binary XOR game corresponding to each of our crietria}\label{appgames}

We merge our notations with the notation of Miller and Shi \cite{millershi}: in the (2,2,2) scenario, a binary nonlocal XOR game is defined by the figure of merit $\sum_{(x,y)\in\{0,1\}^2}f_{xy} E_{xy}$. Miller and Shi have proved conditions for such a game to be self-testing: in particular, the maximum value of the figure of merit must be achieved by a unique probability point up to suitable symmetries.

We want to determine the coefficients $f_{xy}$ for the game that corresponds to our self-testing criterion $E_{xy}=\cos\alpha_{xy}$ with the usual relation \eqref{alphas} for the $\alpha_{xy}$. Before continuing, let us stress that nothing is gained by this translation into a game, and maybe something is lost: indeed, we replace the detailed knowledge of each $E_{xy}$, themselves a function of the collected statistics $p(a,b|x,y)$, with a single number. Nevertheless, the XOR games defined below may be of interest in their own right.

With the geometry of the quantum set, it is easy to guess that a suitable game will be defined by taking
$[f_{00}, f_{01}, f_{10}, f_{11}]$ as the normal $\vec{n}$ to the plane tangent to the quantum set in the desired correlation point. As usual, if the surface of the quantum set is parametrized by $
Q=[x,y,z,g(x,y,z)]$, then $\vec{n}\propto [-\partial_x g, -\partial_y g, -\partial_z g, 1]$ where the partial derivatives are evaluated at the point. Now, the parametrisation of the sector of $Q$ defined by condition \eqref{alphas} is given by the suitable equality in \eqref{t5}, that is
\begin{equation}
\begin{split}
&g(x,y,z) = \sin(\arcsin x + \arcsin y + \arcsin z - \pi)\\&= -\sin(\arcsin x + \arcsin y + \arcsin z)\,.\\
\end{split}
\end{equation}
A possible way of writing $\vec{n}$, in the case where none of the $E_{xy}$ is zero, is
\begin{equation}
\begin{split}
&\left(\begin{array}{c} f_{00}\\ f_{01}\\ f_{10}\\ f_{11} \end{array}\right)\,=\,\left(\begin{array}{c} \sin^{-1}\alpha_{00}\\
-\sin^{-1}(\alpha_{00}+\alpha_{10}+\alpha_{11})\\\sin^{-1}\alpha_{10}\\ \sin^{-1}\alpha_{11} \end{array}\right)\,.\\
\end{split}
\end{equation} As a consistency check, in the case $\alpha_{00}=\alpha_{10}=\alpha_{11}=\pi/4$ we find $f_{xy}=(-1)^{(x+1)y}\sqrt{2}$ which is the CHSH form we used in the main text, up to a multiplicative constant.

\end{document}